\documentclass[11pt]{article}
\usepackage[margin=1.25in]{geometry}
\usepackage[utf8]{inputenc}
\usepackage{authblk}
\usepackage{titling}
\usepackage{amsmath, amssymb}
\usepackage{multirow}
\usepackage{graphicx}
\usepackage[export]{adjustbox}
\usepackage{physics}
\usepackage{hyperref}
\hypersetup{
    colorlinks=true,
    linkcolor=blue,
    filecolor=magenta,      
    urlcolor=cyan,
    }
\begin{titlepage}
\title{\textbf{The No-Short Hair Theorem for Black Holes and Wormholes in Extra-Dimension}}
\author{Mandas Biswas}
\affil{School of Physical Sciences, Indian Association for the Cultivation of Science, Kolkata, India}
\date{January 5, 2026}
\end{titlepage}
\begin{document}
\begin{titlingpage}
    \maketitle
    \begin{abstract}
        The no-short hair theorem for static spherically symmetric black holes in general theory of relativity asserts that if a black hole has hair, that hair must extend beyond the lowest photon sphere radius of the black hole. This report generalizes the theorem to various extra-dimensional cases of black holes and wormholes in post-Einstein gravity and shows how the observational satisfaction of the no-short hair theorem may lead to a probe for the existence of extra-dimensions. We also show how we may lead to argue that this no-short hair theorem is satisfied for black holes in theories of quantum gravity that naturally precludes extra-dimensional cases, like string theory. 
    \end{abstract}
\end{titlingpage}
\newpage
\tableofcontents
\section{Introduction}
Black holes are a class of solutions of Einstein field equations which have an event horizon (a boundary beyond which causality gets violated with respect to us - the observer universe), and region of extreme gravity. For most astrophysical objects, this event horizon lies inside the body - however, under gravitational collapse such bodies may be formed (viz., Chandrashekar limit violation and ECOs) along with possible resolution of horizonless massive objects that mimicks black holes (\cite{MeliaFalcke2001}). Some indirect observations of them through GW QNMs and shadow/lensings are interesting offshoots have led us to verify the conjectured quantum theories of gravity through their data outlook (\cite{FalckeMeliaAgol2000}). One of the central paradox of the quantum nature of gravity in the case of black hole is the information paradox - if everything that goes inside it never comes out and everything eventually evaporates through Hawking radiation (\cite{Hawking1975}), it violates the holy principles of the second law of thermodynamics and quantum mechanical unitarity. That black holes do not contain classical hair is well-established observationally (\cite{IsiEtAl2019}). However, these observations are done at a much lower energy scale to even probe quantum gravitational effects macroscopically near a BH horizon. In such a case, a resolution of the information paradox results into conjecturing "soft hair" (\cite{HawkingPerryStrominger2016}) - which is thought to carry information about particles that have fallen into the black hole. The soft hair arises from the quantum excitations of these soft particles on the black hole's event horizon, as is believed. This hair must be visible through non-trivial observational bounds of quantum GW QNM analysis. 
\\That static, spherically symmetric black holes are canonical, that is, they are specified completely by the conserved Noether charges defined at infinity (mass - from energy conservation (Schwarzschild), electric charge - from U(1) global gauge symmetry conservation (Reissener-Nordstrom), and angular momentum - from rotational symmetry conservation (Kerr)) came to be known as the no-hair theorem for these black holes, as conjectured in the 1970s \cite{bekenstein1998blackholesclassicalproperties}. They also hold good for certain other classes of black holes as well as explicitly shown in several papers over the 1970s and 1990s (\cite{Bekenstein1972PRL,Bekenstein1972PRD1,Bekenstein1972PRD2,Teitelboim1972,Heusler1992}), involving various scalar and vector fields - such that the baryon and lepton number conservation is kind of unobservable in the case of black holes, hence, "transceded"; along with that, coupling between mesons and the black holes becoming zero at the horizon beginning from the central singularity. 
\\However, some theories like Einstein-Skyrme and Einstein NA-Proca seems to have hairy black hole stable linearly perturbative solution and hence dodges this theorem (\cite{Bizon1994}).
\\It is clear by now that by hair we mean the charges that one requires to specify a black hole solution other than the three canonical charges. An interesting argument in support of calling this hair is the fact that the matter content of the theories that we mentioned above has a non-linear behavior - it seems that the interaction between the part of the matter field which would be radiated away and which would be sucked inside the black hole is the cause of the possession of hair by black holes. This would also imply that this non-linear behavior have to be present both very close (where the field would tend to be sucked in) and away enough from the horizon (where the field would tend to be radiated away), with the self-interacting binding the fields in these two regions together. In fact for a scalar field with a convex potential (like the mass term in Klein-Gordon field) this proof is extremely simple and does not even require Einstein's equation to prove that there is no hair (\cite{Bekenstein1995}).   
\\We use the term "Hairosphere" to refer to the loosely defined region where the non–linear behavior of the fields is present, in contrast to the asymptotic region where the behavior of the fields is dominated by linear terms in their respective equations of motion. A slightly more explicit characterization of this region is given by the region where $r^4T^r_r$ does not tend to zero locally at infinity. However, this characterization is not required in higher dimensions (from $d\geq5$) as $r^dT^r_r\longrightarrow0$ even in higher-dimensional RN BHs.  
\\That there might be extra dimensions to our world in compactified form was first proposed in 1921\cite{Kaluza1921}. Low energy limits of quantum gravity theories in higher-dimensional string theories (a fully quantum theory of gravity) provide various modified gravity theories effectively in higher-dimensions which have found footing in the recent years in their ability to address anomalous behavior of gravitational phenomena within a cosmological setting (Einstein gravity + cosmological constant) - a natural selection to bypass several issues that result from dark matter/dark energy in asymptotically flat spacetime. Although perhaps inaccessible from a physical point of view, additional spatial dimensions might help augment the range of stability for matter under extreme conditions where higher-order curvature effects become important. (\cite{Farnes2018})
\\This article deals with bounds of the hair behavior (regardless of classical or quantum hair, that is, "hard" or "soft" hair\cite{Coleman_1992}), and thus, may shed light upon the behavior of this information carrying hairs hinting at a connection to black hole entropy in more exotic theories of quantum gravity. However, the article - even though deals with modified theories of gravity which may arise as classical limits of different quantum theories of gravity like holography or superstring theory - does not deal with quantum gravity explicitly. We will try to understand how the no-short hair theorem might be generalized for extra-dimensional theories of gravity which accommodate solutions of static spherically symmetric black holes. We will first start with d-dimensional general relativity, then move on to first order Lovelock generalizations of general relativity - that is, 5D EGB theory of gravity. We will also try to see how this no-short hair theorem will hold for Randall-Sundrum braneworld scenario too - where our universe is considered a 5D AdS bulk with our 4D world living on Planck 3-brane with positive tension, with possibility of a negative-tension brane being present or not inside the orbifold. We will also generalize the theorem for certain models of wormholes as well, and show how extra-dimensional effects can probe braneworld wormholes to have no-short hair behavior compared to isolated galactic wormholes. Alongside that, we will show how Brans-Dicke theory of gravity in 4D probes for no-short hair theorem for black holes, where the details of the calculation is given at Appendix A. In Appendix B, we use the null circular orbit expressions of a general static spherically symmetric black hole and derive a theory-agnostic form of the no-short hair theorem for black holes. 
\\This work is heavily indebted to \cite{NunezQuevedoSudarsky1996}.
\section{Review of the No-Short Hair Theorem in 4D Einstein Gravity}
The ansatz metric is,
\begin{equation}
    ds^2 = -e^{-2\delta}\mu dt^2 + \mu^{-1} dr^2 + r^2 (d\theta^2 + \sin^2\theta d\phi^2)
\end{equation}
where, $\delta$ and $\mu = 1 - \frac{2m(r)}{r}$, are functions of r only, and we assume that there is a regular event horizon at $r=r_H$, so, $m(r_H) = r_H/2$ and $\delta(r_H)$ is finite. Asymptotic flatness requires, in particular, that $\mu\longrightarrow 1$ and $\delta\longrightarrow 0$, at infinity. This metric describes a fairly large class of static black holes in higher dimensions as long as the area gauge is satisfied \cite{HorowitzEtAl2022}.
\\The EM-tensor will have only one non-trivial component in conservation equation,
\begin{equation}
     \nabla_{\mu}T^{\mu}_{r} = 0
\end{equation}
We will take the EM-tensor of a perfect anisotropic fluid distributed in the metric, which is given by, $T_{\mu\nu} = (-\rho, P_r, P, P)$. Relevant Einstein tensors are,
\begin{eqnarray}
    G_{tt} &=& -e^{-2\delta}\mu\frac{-1+\mu+r\mu'}{r^2}\\
    G_{rr} &=& \frac{-1+\mu-2r\mu\delta'+r\mu'}{r^2\mu}
\end{eqnarray}
Einstein field equation is given by,
\begin{equation}
    R_{\mu\nu} - \frac{1}{2}g_{\mu\nu}R = G_{\mu\nu} = 8\pi T_{\mu\nu}
\end{equation}
We get,
\begin{eqnarray}
    \mu' &=& 8\pi rT^{t}_{t} + \frac{1-\mu}{r}\\
    \delta' &=& \frac{4\pi r}{\mu}(T^{t}_{t} - T^{r}_{r})
\end{eqnarray}
We begin with,
\begin{equation}
    \nabla_{\mu}T^{\mu}_{r} = \partial_\mu T^{\mu}_{r} + \Gamma^{\mu}_{\mu\sigma}T^{\sigma}_r - \Gamma^{\sigma}_{\mu r}T^{\mu}_{\sigma}
\end{equation}
Substituting the Christoffel symbols, and upon algebraic manipulation with them, we will arrive at,
\begin{equation}
    e^\delta(e^{-\delta}r^4T^r_r)' = \frac{r^3}{2\mu}[(3\mu - 1)(T^r_r - T^t_t) + 2\mu T]
\end{equation}
\subsection{The Theorem}
\textbf{Theorem:} Let equation (1) represent the line element of an asymptotically flat static spherically symmetric black hole space time, satisfying Einstein’s equations, with matter
fields satisfying the WEC and such that the trace of the energy momentum tensor is non-positive, and such that the energy density $\rho$ goes to zero faster than $r^{-4}$, then the function
$\varepsilon = e^{-\delta}r^4T^r_r$ is negative semi-definite at the horizon and is decreasing between $r_H$ and $r_0$ where $r_0 > \frac{3}{2}r_H$ and from some $r > r_0$, the function $\varepsilon$ begins to increase towards its asymptotic value, namely 0.
\\\textbf{Proof:} For a step, we go to the proper radial distance $dx = \mu^{-1/2}dr$ to get rid of the coordinate singularity at $r=r_H$. We rewrite the equation (9) as,
\begin{equation}
    \frac{d(e^{-\delta}r^4T^r_r)}{dx} = \frac{e^{-\delta}r^3}{2}[\mu^{1/2}(3(T^r_r - T^t_t) + 2T) - \mu^{-1/2}(T^r_r - T^t_t)]
\end{equation}
Since the components $T^r_r, T^\theta_\theta, \text{and } T^\phi_\phi$ must be regular at the horizon (i.e, the scalar $T^{\mu\nu}T_{\mu\nu}$is regular at the horizon) and, since x is a good coordinate at the horizon (true for cosmological background BHs too; another way to see this is to define a normal null vector to the horizon and take its norm with the EM-tensor - that must be regular at the horizon as well), the LHS of the equation must be finite, and hence, to make sure that $\mu^{-1/2}$ does not blow up as $\mu\longrightarrow 0$, we have,
\begin{equation}
    T^r_r(r_H) = T^t_t(r_H) = -\rho 
\end{equation}
This implies that $\varepsilon(r_H) \leq 0$. Now, as $T<0$, and $(T^r_r - T^t_t)>0$ (by the WEC), we have, from the RHS of equation (9), that it is negative definite, unless, $3\mu-1>0$. Thus, $\varepsilon$ is a decreasing function of r, at least upto the point until $3\mu-1$ becomes positive, and this occurs at $r_1 = 3m(r_1)$, by simple inequality algebra,
\begin{eqnarray}
    3\mu - 1 &>& 0\nonumber\\
    \implies \mu &>& \frac{1}{3}\nonumber\\
    \implies 1 - \frac{2m}{r} &>& \frac{1}{3}\nonumber\\
    \implies m &<& \frac{r}{3}
\end{eqnarray}
which implies $r_0>r_1$. Since, $m(r)$ is an increasing function of r (as follows from the WEC and that equation (7) can be written as $m' = 4\pi r^2\rho$, by putting $\mu = 1 - \frac{2m(r)}{r}$, and differentiating both sides with respect to r), we then have, 
\begin{equation}
    r_0 > 3m(r_1) > 3m(r_H) = \frac{3}{2}r_H
\end{equation}
where the last part of the inequality follows from the fact that $r_H$ must be lesser than that of the point of positivity of $3\mu-1$.
\\That completes the proof. (Q.E.D.)
\\This theorem is applicable to a wide range of BH solutions in GR (both singular and regular) on which the usual no-hair theorem is not applicable. Along with this, a physical argument regarding the surface gravity of the black hole measured at the $m(r_H)$ is also provieded in \cite{AuthorMastersThesis}
\section{d-Dimensional Einstein Gravity}
The action is given by,
\begin{equation}
    S = \int d^dx(\frac{1}{2k}\sqrt{-g}R + L_m)
\end{equation}
Inside this $k$, we will get a Newton's constant of d-dimensions, given by, 
\begin{equation}
    G_d = G\frac{d-2}{d-3}A_{d-2};\text{ }A_{d-2} = \frac{2\pi^{\frac{d-1}{2}}}{\Gamma[(d-1)/2]}
\end{equation}
The variation of this action by familiar means of 4D will lead to d-dimensional Einstein field equations,
\begin{equation}
    G_{ab} = R_{ab} - \frac{1}{2}g_{ab}R = 8\pi T_{ab}
\end{equation}
The ansatz metric will be the same static spherically symmetric nature, so we will represent the metric in $(t, r, \theta, \phi, \psi, \eta, ...)$, where the latter coordinates are all angular. The metric is, therefore, 
\begin{equation}
    ds^2 = -e^{-2\delta}\mu dt^2 + \mu^{-1} dr^2 + r^2d\Omega^2_{d-2}
\end{equation}
where, every condition remains the same except that, $\mu = \frac{16\pi m(r)}{(d-2)A_{d-2}r^{d-3}}$. Regular horizon implies, $m(r_H) = \frac{d-2}{16\pi}A_{d-2}r_H^{d-3}$. The Christoffel symbols will be easily derived using method of induction, and they can be used to arrive at the hair function.
The d-dimensional necessary Einstein field equations on $\mu$ and $\delta$ will be given by (I found this out by induction used on eqn. (17)),
\begin{eqnarray}
    \mu' &=& \frac{16\pi T^t_t r}{d-2} + \frac{(d-3)(1-\mu)}{r}\\
    \delta' &=& \frac{8\pi r}{(d-2)\mu}(T^t_t - T^r_r)
\end{eqnarray}
Starting from eqn. (2), we will get,
\begin{equation}
    e^\delta(e^{-\delta}r^dT^r_r)' = \frac{r^{d-1}}{2\mu}[(\mu(d-1) + 3 - d)(T^r_r - T^t_t) + 2\mu T]
\end{equation}
The proof of the no-short hair theorem will be exactly analogous until,
\begin{eqnarray}
    \mu(d-1) + 3-d &>& 0\nonumber\\
    \implies \mu &>& \frac{d-3}{3-1}\nonumber\\
    \implies r_0 &>& (\frac{d-1}{2})^{\frac{1}{d-3}}r_H
\end{eqnarray}
The last step is implied from the same argument as the 4D GR case. So, the modified theorem for d-dimensional GR is:
\\\textbf{Theorem:} Let equation (18) represent the line element of an asymptotically flat static spherically symmetric black hole space time, satisfying Einstein’s equations in d-dimensions, with matter fields satisfying the WEC and such that the trace of the energy momentum tensor is non-positive, then the function
$\varepsilon = e^{-\delta}r^dT^r_r$ is negative semi-definite at the horizon and is decreasing between $r_H$ and $r_0$ where $r_0 > (\frac{d-1}{2})^{\frac{1}{d-3}}r_H$ and from some $r > r_0$, the function $\varepsilon$ begins to increase towards its asymptotic value, namely 0.
\\Note that, in $d\geq5$ Maxwell fields go to asymptotical zero faster than $r^d$. Hence, this condition of the theorem may be relaxed in higher-dimensional general relativity - as it is apparent only those matter fields are allowed in those dimensions which automatically satisfy these equations. 
\section{5D EGB Theory}
The Lovelock generalisation of general theory of relativity, allows for extensions to higher-dimensional metrics. A particular case of Lovelock theory, the Einstein–Gauss–Bonnet (EGB) gravity (which is the solution upto second order of Lovelock theory, the zeroth order being the cosmological constant), contains geometrical features that favor a covariant theory of gravity (standard requirements of diffeomorphism and Bianchi identities are satisfied de facto)\cite{Padmanabhan_2013, PhysRevD.102.084028}. The Lagrangian in EGB theory includes the Gauss–Bonnet term which is quadratic in the Riemann tensor; however, second-order quasilinear equations of motion are obtained and so no additional dynamical degrees of freedom are introduced as in f(R) gravity. The quadratic nature of the Gauss–Bonnet term is consistent with the low energy effective action in heterotic string theory\cite{Polchinski1998V2}. This has contributed to Lovelock theory having acquired a reputation of being a string-theory-inspired gravity theory. The Lanczos-Lovelock gravity is special in the sense that the field equations derived from the Lanczos-Lovelock action contain only second order derivatives of the metric and have natural thermodynamic interpretation. \cite{BoulwareDeser1985} found solutions to vacuum equations that satisfy the EGB field equations in 5D and 6D. 
\\The action of the Gauss-Bonnet gravity is given by,
\begin{equation}
    S = \frac{1}{2k}\int\sqrt{-g}(R + \alpha L_{GB})d^5x + S_{matter}
\end{equation}
$\alpha$ is related to string tension in heterotic string theory. It was determined that the energy density due to the Gauss–Bonnet term should not exceed 15\%. Stronger constraints have been obtained more recently from gravitational wave observational data \cite{LyuJiangYagi2022}. For this case only, I will let the Greek indices run from 0 to 4. The metric signature is (- + + + +). 
\\Usual variation of the action using the identities derived and used earlier will lead to the EGB field equations,
\begin{equation}
    G_{\mu\nu} + \alpha H_{\mu\nu} = 3\pi^2T_{\mu\nu}
\end{equation}
$G_{\mu\nu}$ is the usual Einstein tensor in 5D, and the Lanczos tensor is given by,
\begin{equation}
    H_{\mu\nu} = 2(RR_{\mu\nu} - 2R_{\mu\rho}R^\rho_\nu - 2R^{\rho\sigma}R_{\mu\nu\rho\sigma} + R^{\rho\sigma\lambda}_\mu R_{\nu\rho\sigma\lambda}) - \frac{1}{2}g_{\mu\nu}L_{GB}
\end{equation}
We defined the Lovelock tensor to be,
\begin{equation}
    L_{GB} = R^2 - 4R_{\mu\nu}R^{\mu\nu} + R_{\mu\nu\rho\sigma}R^{\mu\nu\rho\sigma}
\end{equation}
Our ansatz metric here will be,
\begin{equation}
    ds^2 = -e^{-2\delta}f dt^2 + f^{-1}dr^2 + r^2d\theta^2 + r^2sin^2\theta d\phi^2 + r^2sin^2\theta sin^2\phi d\psi^2
\end{equation}
The assumptions and conditions on this will remain same as the 4D GR case, and the WEC will hold, implying, for a shear-free pressure-isotropic stellar fluid present around the black hole,
\begin{equation}
    T^\mu_\nu = \text{diag}(-\rho, p_r, p, p, p)
\end{equation}
The $\mu$ in 4D GR becomes f(r) here, with the expression, $f(r) = 1 + \frac{r^2}{2\alpha}(1 - \sqrt{1 + \frac{8\alpha\mu(r)}{3r^4}})$. Here, $\mu(r)$ is related to the mass function by a constant, given by, $m(r) = \frac{\mu(r)\times 2\pi^2}{8\pi G} = \frac{\pi\mu(r)}{4G}$. The horizon solves the equation, $r_H = \sqrt{\frac{2\mu(r_H)}{3} - \alpha}$, such that, $f(r_H) = 0$. Note that in the $\alpha \longrightarrow 0$ limit, the black hole solution reduces to the Einstein-Tangherlini one with varibale mass function,
\begin{equation}
    f(r) = 1 - \frac{2\mu(r)}{3r^2} + O(\alpha) +...
\end{equation}
Note, however, that some solutions to the EGB equations are found to diverge as $\alpha\longrightarrow0$, including the structure we have taken. \cite{BeroizDottiGleiser2007}
\\Varying the action against the metric then generates the Einstein–Gauss–Bonnet equations which follow below,
\begin{eqnarray}
    \pi^2\rho &=& -\frac{f}{r^2}[1 - \frac{1}{f} + \frac{f'}{2fr}(r^2 + 4\alpha(1 - f))]\nonumber\\
    \implies \frac{f'}{2} &=& \frac{\pi^2 T^t_t - \frac{f}{r^2} + \frac{1}{r^2}}{F(r)}\\
    \pi^2 p &=& \frac{f}{r^2}[1 - \frac{1}{f} + (\frac{f'}{2fr} - \frac{\delta'}{r})(r^2 + 4\alpha(1-f))]\nonumber\\
    \implies -\delta' &=& \frac{\pi^2(T^r_r - T^t_t)}{fF(r)}
\end{eqnarray}
where, $F(r) = \frac{1}{r} + \frac{4\alpha(1-f)}{r^3}$, which I used for brevity. 
\\We again start from the Bianchi identity,
\begin{equation}
    \nabla_{\mu}G^{\mu}_r = \nabla_{\mu}T^{\mu}_r = 0
\end{equation}
We then get,
\begin{equation}
    \partial_rT^r_r + \frac{3}{r}T^r_r - \delta'T^r_r + \delta'T^t_t + \frac{1}{2}\frac{f'}{f}(T^r_r - T^t_t) - (T^\theta_\theta + T^\phi_\phi + T^\psi_\psi) = 0
\end{equation}
Rearranging the terms as per the algorithm of 4D GR, and getting a hint from the structure of the hair function of the d-dimensional GR case, we get the hair function expression for 5D EGB theory,
\begin{equation}
    e^\delta(e^{-\delta}r^5T^r_r)' = \frac{r^4}{f F(r)}[(T^r_r - T^t_t)(f F(r) + (f-1)r^{-1}) - f F(r)T]
\end{equation}
It is clear like the argument in the 4D GR theory, that, $T^t_t(r_H) = T^r_r(r_H) = -\rho(r_H) < 0$, so, the hair function is negative at the horizon of the EGB black hole. Everything in the proof ongoing is the same, upto the point where,
\begin{eqnarray}
    f F(r) + \frac{f - 1}{r} &>& 0\nonumber\\
    \implies 4\alpha f^2 - f(2r^2 + 4\alpha) + r^2 &<& 0\nonumber
\end{eqnarray}
So, we get, $f(r)$ must lie within the range, 
\begin{equation}
    \frac{r^2}{4\alpha} + \frac{1}{2} - \frac{1}{4}\sqrt{\frac{r^4}{\alpha^2} + 4} < f(r) < \frac{r^2}{4\alpha} + \frac{1}{2} + \frac{1}{4}\sqrt{\frac{r^4}{\alpha^2} + 4}\nonumber
\end{equation}
Now, since the spacetime is asymptotically flat, by matching the boundary condition, we can say that,
\begin{equation}
    f(r) > \frac{r^2}{4\alpha} + \frac{1}{2} - \frac{1}{4}\sqrt{\frac{r^4}{\alpha^2} + 4} = f(r_1)
\end{equation}
with the other solution, becoming physically unfeasible. 
The photon sphere of this metric also solves the equation (106), and upon solving that (with $\mu$ now being a constant, as in Schwarzschild case), we get,
\begin{equation}
    r_{ph} = \sqrt[4]{\frac{16\mu^2 - 24\alpha\mu}{9}}
\end{equation}
We assume, $\mu(r)$ must be a positive function for all $r$, because we are only dealing with positive mass. For the same reason, $f(r) > 0$ for the metric outside the horizon. For the proof to hold the inequality after (34), we have,
\begin{eqnarray}
    F(r) &>& 0\nonumber\\
    \implies f &<& \frac{r^2}{4\alpha} + 1\nonumber\\
    \implies \frac{8\alpha\mu(r)}{3r^4} &>& -\frac{3}{4}
\end{eqnarray}
It is consistent with the fact that $\alpha > 0$ is the only condition upon which there is no constraint on the expression of $\mu(r)$ and the inequality is satisfied. 
\\So, $\alpha > 0$ is the \textbf{only condition upon which the sufficiency} of the no-short hair theorem lies in 5D EGB theories. 
\\To find the relation between lower bound of $f(r)$ and the photon sphere radius of (35), we have to find out what the expression of $f'$ tells us from the field equation (29). Upon visiting that, and simplifying it in a straightforward but tedious procedure, we get,
\begin{equation}
    \frac{r}{4}\frac{1}{g(r)}(\frac{8\mu'}{3r^4}) = -\frac{\pi^2T^t_t}{F(r)} + (1-g(r))(\frac{1}{2\alpha F(r)} + \frac{1}{\alpha}) + \frac{8\mu(r)}{g(r)3r^4}
\end{equation}
where, $g(r) = \sqrt{1 + \frac{8\alpha\mu}{3r^4}} > 1$. Now, it is clear that the first and third term on the RHS is positive, and the second term is negative. The negativity comes from the first factor of the second term. Now, for all the theories that were calculated the energy-momentum tensor in the 4D GR case, all stable black hole solutions of the EFE - we found out that $T^t_t \sim \frac{1}{r^4}$ at the minimum\cite{AuthorMastersThesis}. Now, since $T^t_t$ is the component which contains the information of the $\mu(r)$, we can safely assume that $\mu(r)$ falls off faster than $1/r^5$. In such a case, the term within the square root of $g(r)$ will approach to one very fast as we move away from the horizon, and the horizon itself is dependent upon the fall of $\mu(r)$ - and so, we can conclude that the contribution of the second term on the RHS of (37) is lesser than the positive ones. In particular, if we look at the near-Schwarzschild expansion of (1 - g(r)), we have,
\begin{equation}
    1 - g(r) \approx 1 - 1 - \frac{1}{2}\frac{8\alpha\mu}{3r^4} + O(\alpha^2) = -\frac{4\alpha\mu}{3r^4}; \text{ }F(r)^{-1}\text{, }g(r)^{-1} \approx 1\nonumber
\end{equation}
Thus, upto the first order of $\alpha$, we find,
\begin{equation}
    \frac{r}{4}(\frac{8\mu'}{3r^4}) \approx -\pi^2T^t_t - \frac{4\mu}{3r^4}(\frac{1}{2} + 1) + \frac{8\mu}{3r^4} > 0
\end{equation}
Thus, $\mu$ is an increasing function of r at a distance considerably away from the horizon, which is what we are concerned about. So, using $\mu(r_1) > \mu (r_H)$, we can find a more refined bound of the "hairiness", which might be suitable for probing black holes in effective quantum gravity.\cite{liu2024light}
\section{BH Mimicking and Traversable Wormholes in 4D}
This wormhole is a special solution of 4D ordinary general theory of relativity. They connect two distinct universes via a throat. They are conjectured as a viable protocol for quantum teleportation if they are stable, unlike Einstein-Rosen bridge. This stability may give rise to exotic matter content (violating various energy conditions) that gives rise to wormhole solutions. In fact, if the two universes have non-identical photon spheres with each other, one's angular momentum barrier acts as a partial reflecting surface to the another, giving a pseudo-horizon type of behavior, thus mimicking BHs like ECOs. We are yet to observationally detect wormholes in our universe. 
\subsection{Damour-Solodukhin Wormholes}
The metric is given by (which is in a special form of our original ansatz metric for 4D Einstein gravity),\cite{Damour_2007}
\begin{equation}
    ds^2 = -(1 - \frac{2M_1}{r})dt^2 + (1 - \frac{2M_2}{r})^{-1}dr^2 + r^2d\Omega_2^2;\text{ }M_2 = M_1(1+\lambda^2)
\end{equation}
As is clearly evident, $r = 2M_2$ is a null hypersurface, but it is not the Killing horizon of this geometry, as the norm of the Killing vector field $\partial_t$ for this metric is given by $g_{tt}$, which is pretty well defined at $r=2M_2$. 
\\The calculation of the Kretschmann scalar shows us that, $R^{\mu\nu\alpha\beta}R_{\mu\nu\alpha\beta} \propto r^{-6}(r-2M_1)^{-4}$, which diverges at $r+2M_1$. So, there is a physical singularity at $r+2M_1$, but well-defined at $r=2M_2$. Therefore, if one continues the spacetime through $r=2M_2$, one gets a naked singularity at $r=2M_1$. To avoid that, we say that this is a wormhole geometry with the null throat at $r=2M_2$. 
\\This is the solution of the Einstein field equations iff the RHS is given by the following anisotropic fluid components (for isolated DS wormholes),
\begin{equation}
    T^{\mu(w)}_\nu = \text{diag.}(0, p_r^{(ds)}, p_p^{(ds)}, p_p^{(ds)});\text{ }p_r^{(ds)} = -\frac{\lambda^2M_1}{4\pi r^2(r-2M_1)};\text{ }p_p^{(ds)} = \frac{(r-M_1)\lambda^2M_1}{8\pi r^2(r - 2M_1)^2}
\end{equation}
So, the geometrical conditions of the theorem is identically satisfied. Following \cite{Biswas_2024} we can assume that dark matter content can be added on top of this anisotropic fluid for a true galactic wormhole which behaves as ordinary matter satisfying WEC by itself, $T^\mu_\nu = (-\rho, p, p, p)$. A specific form is used in the last paper where they show that WEC gets violated in presence of this dark matter halo only very near the wormhole throat, but as we move away - it is uniformly satisfied. The three other conditions of the theorem, if they have to be satisfied, the following conditions has to hold,
\begin{itemize}
    \item The trace of the EM-tensor must be less than zero. Since the EM-tensor of the galactic dark matter is already present (the perfect fluid EM-tensor, like the case of ordinary 4D GR), we will need to consider the contribution of both these in the expression of trace. That will imply,
    \begin{equation}
        r > M_1 + \sqrt{M_1^2 + \frac{\lambda M_1}{2\sqrt{\pi(\rho-3p)}}} = r_1\text{ (say)}
    \end{equation}
    \item Now, it has to be outside the null hypersurface, that is, the value of $\lambda$ should be such that,
    \begin{eqnarray}
        r &>& 2M_2\nonumber\\
        \implies 4\lambda(1 + \lambda^2) &<& \frac{1}{2M_1\sqrt{\pi(\rho-3p)}}
    \end{eqnarray}
    \item The WEC must hold. Since, the wormhole matter does not satisfy the WEC, and the hair behavior must extend beyond the point where $T^r_r$ must show its non-linear behavior which is contained in the galactic matter EM-tensor, we get a condition,
    \begin{equation}
        |p_r^{(ds)}(r_1)|<<p(r_1);\text{ }p_p^{(ds)}(r_1)<<p(r_1);\text{ }r_0>r_1
    \end{equation}
\end{itemize}
With these conditions satisfied, the exact same no-short hair theorem holds for DS-wormhole too, as was the case in black holes.
\\As a result, we can see that isolated Damour-Solodukhin wormholes won't show no-short hair behavior. 
\subsection{Morris-Thorne Wormholes}
The metric for a static, traversable wormhole \cite{MorrisThorne1988} is given by,
\begin{equation}
ds^{2} = - e^{2\Phi(r)} \, dt^{2} + \frac{dr^{2}}{1 - \dfrac{b(r)}{r}} + r^{2}\left( d\theta^{2} + \sin^{2}\theta \, d\phi^{2} \right)
\end{equation}
where, $\Phi(r)$ is the redshift function, which is demanded to be finite everywhere and hence there is no horizon, thus being traversable; and $b(r)$ is the shape function which determines the spatial distribution of the wormhole, and the throat is at $b(r_0) = r_0$. We also have the flare-out condition, $b'(r_0)<1$, which results in violating the NEC (which is a weaker condition than WEC) and asymptotic flatness requires $\Phi(r)\longrightarrow0$ and $b(r)/r \longrightarrow0$ as $r\longrightarrow\infty$. 
\\We can define a proper radial distance and rewrite the metric in those terms given by,
\begin{eqnarray}
l(r) &=& \pm \int_{r_0}^{r} \frac{dr}{\sqrt{1 - \dfrac{b(r)}{r}}}\text{ (Denoting both sides of the geometry)}\nonumber\\
\implies ds^{2} &=& - e^{2\Phi(l)} \, dt^{2} + dl^{2} + r^{2}(l)\, d\Omega^{2}
\end{eqnarray}
Using EFE in 4D, we can find,
\begin{eqnarray}
    \rho(r) = \frac{1}{8\pi}\frac{b'(r)}{r^{2}}\nonumber\\
    p_{r}(r) = \frac{1}{8\pi} \left[- \frac{b(r)}{r^{3}}+\frac{2}{r}\left(1 - \frac{b(r)}{r}\right)\Phi'(r)\right]
\end{eqnarray}
At the throat, we will have,
\begin{equation}
   \left. \left( \rho + p_{r} \right) \right|_{r_0} = \frac{1}{8\pi r_{0}^{2}}\left[b'(r_{0}) - 1\right]
\end{equation}
The flare-out condition makes sure that $\rho+p_r<0$, so NEC, and eventually WEC gets violated (however, to be noted, that in principle, $\rho$ can be positive at the throat by equation (46)). Thus, traversable wormholes will not obey the no-short hair theorem in classical general relativity. There have been some studies in constructing NEC-obeying (but WEC disobeying) traversable wormhole solutions in certain modified gravity theories, see \cite{Baruah_2019}. In particular, we look at traversable wormhole solutions in EGB theory, where $p_r + \rho = (d - 2)\left[ 1 + 2 \bar{\alpha} Q \right]\left( N - P \right)$, where, $N = - \frac{\Phi'}{r}\left( 1 - \frac{b}{r} \right)$, $P = \frac{1}{2 r^{3}}\left( b' r - b \right)$ and $\alpha'$ is related to be Gauss-Bonnet constant by a linear algebraic equation. It can easily be shown by evaluating this expression at the throat that if $\alpha>0$, NEC and WEC will always be violated. It was shown that for 6D,
exotic mater can be limited to an arbitrarily small region, much like in the case of galactic DS wormhole. AdS wormholes have been found that satisfy $\alpha>0$ and still have normal matter everywhere throughout the spacetime.\cite{MaedaNozawa2008} 
\section{Braneworld Black Holes in 5D Bulk}
Braneworld model is a theory which describes our universe in terms of warped extra-dimensional space geometry. It was first proposed by Randall and Sundrum who wanted to address the mass hierarchy problem. According to this theory, our world is actually a 5d AdS space where there are two branes of (1+3)-dimensions where all the elementary particles reside, compactified, except the graviton - which can propagate freely in the extra-dimension. The location of the branes differentiate the two RS models - if the branes are placed at a finite distance to themselves, it is RS1 (and our 4d universe is the brane with the negative tension; the positive tension brane being the "Planck brane")\cite{RandallSundrum1999RS1}, while the RS2 model has a brane infinitely far away from the first\cite{RandallSundrum1999RS2}. The brane tension arises from the extremely warped nature of spacetime.
\\The canonical form of the RS metric without any black hole is given by (in both RS1 and RS2),
\begin{equation}
    ds^2 = \frac{1}{k^2y^2}(dy^2 + \eta_{\mu\nu}dx^\mu  dx^\nu)
\end{equation}
where, k is some constant. The space has boundaries at $1/k$ and $1/Wk$, where W is the warp factor. $y=1/k$ is the Planck brane, and $y=1/Wk$ is the TeV brane - and our universe resides on the TeV brane. The distance between the two branes is given by, $\frac{\ln(W)}{k}$. 
\\We are concerned with the BBH solutions. For that, we require the geometry of the 4d brane with respect to the bulk 5D. In the induced metric of the brane from the bulk, we can have the Gauss and Codazzi equations (in the 5d metric $g_{\mu\nu}$, and the 4d induced metric is given by $q_{\mu\nu} = g_{\mu\nu} - n_\mu n_\nu$, where, n is the normal vector of the domain wall brane towards the bulk)\cite{Whisker2008},
\begin{eqnarray}
    ^{(4)}R^{\alpha}{}_{\beta\gamma\delta} &=& {}^{(5)}R^{\mu}{}_{\nu\rho\sigma} \, q^{\alpha}{}_{\mu} \, q^{\nu}{}_{\beta} \, q^{\rho}{}_{\gamma} \, q^{\sigma}{}_{\delta} + K^{\alpha}{}_{\gamma} K_{\beta\delta} - K^{\alpha}{}_{\delta} K_{\beta\gamma}\nonumber\\
    D_{\nu} K^{\nu}{}_{\mu} - D_{\mu} K &=& {}^{(5)}R_{\rho\sigma} n^{\sigma} q^{\rho}{}_{\mu}
\end{eqnarray}
where, the extrinsic curvature\footnote{The curvature which comes due to embedding of the domain wall in the surrounding space. This can be better understood by trajectories of the closed geodesics on the space internally.} of the brane is denoted by $K_{\mu\nu} = q^{\alpha}{}_{\mu} \, q^{\beta}{}_{\nu} \, \nabla_{\alpha} n_{\beta}$, where its trace is given by $K = K^{\mu}{}_{\mu}$ and $D_{\mu}$ is the covariant derivative compatible with $q_{\mu\nu}$.
\\Contracting the Gauss-Codazzi equations, we can get the $
^{(4)}G_{\mu\nu}$, and then using the 5D Einstein equation with a source and the 5D decomposition of the Riemann tensor into the Weyl curvature, Ricci tensor, and the Ricci scalar, we obtain the effective 4D equation on the brane domain wall, given by,
\begin{eqnarray}
    ^{(4)}G_{\mu\nu} &=& \frac{16\pi}{3} \left[ T_{\rho\sigma} q^{\rho}{}_{\mu} q^{\sigma}{}_{\nu} + \left( T_{\rho\sigma} n^{\rho} n^{\sigma} - \frac{1}{4} T^{\rho}{}_{\rho} \right) q_{\mu\nu} \right] + K K_{\mu\nu} - K^{\sigma}{}_{\mu} K_{\nu\sigma} \nonumber\\
    &-& \frac{1}{2} q_{\mu\nu} \left( K^2 - K^{\alpha\beta} K_{\alpha\beta} \right) - E_{\mu\nu}
\end{eqnarray}
where, $E_{\mu\nu} = ^{(5)}C^{\alpha}_{\beta\rho\sigma}n_\alpha n^\rho q^\beta_\mu q^\sigma_\nu$. This is the electric part of the Weyl tensor. Using the Codazzi equation, and the 5D Einstein equation, we will get, $D_{\nu} K^{\nu}{}_{\mu} - D_{\mu} K = 8\pi \, T_{\rho\sigma} \, n^{\sigma} q^{\rho}{}_{\mu}$. Along with this, we choose a particular coordinate system with respect to the normal, such that, $n_{\mu} dx^{\mu} = dy$, such that it constrains the extra-dimension acceleration of the coordinate $a^{\mu} = n^{\nu} \nabla_{\nu} n^{\mu} = 0$. The metric, thus becomes, 
\begin{equation}
    ds^2 = dy^2 + q_{\mu\nu} dx^{\mu} dx^{\nu}
\end{equation}
Bearing the bulk structure in mind, we can decompose the 5D EM-tensor such that, $T_{\mu\nu} = -\Lambda g_{\mu\nu} + S_{\mu\nu} \, \delta(y)$, with the surface EM-tensor on the domain wall given by, $S_{\mu\nu} = -\lambda q_{\mu\nu} + \tau_{\mu\nu}$. With $\tau_{\mu\nu}n^\mu = 0$. \(\Lambda\) is the cosmological constant of the bulk spacetime. \(\lambda\) and \(\tau_{\mu\nu}\) are the vacuum energy and the energy-momentum tensor, respectively, in the brane world. Note that \(\lambda\) is the tension of the brane in five dimensions. Israel's junction condition \cite{Israel1966} ensures that the extrinsic curvature is uniquely determined by the surface EM-tensor on the domain wall, imposed with the $\mathbb{Z}_2$-symmetry - both sides of the brane carries the same dynamics of the bulk. For completeness, we enlist the equation here,
\begin{equation}
    \lim_{y\longrightarrow0}K_{\mu\nu} = -8\pi(S_{\mu\nu} - \frac{1}{3}q_{\mu\nu}S)\nonumber
\end{equation}
Thus substituting the last equation in equation (46), we get, the effective 4D Einstein equation on the brane, such that,
\begin{equation}
    ^{(4)}G_{\mu\nu} = -\Lambda_4 q_{\mu\nu} + 8\pi G_N \tau_{\mu\nu} + \kappa_5^4 \pi_{\mu\nu} - E_{\mu\nu}
\end{equation}
where, 
\begin{eqnarray}
    \Lambda_4 &=& \frac{1}{2} 64\pi^2 \left( \Lambda + \frac{1}{6} 64\pi^2 \lambda^2 \right)\nonumber\\
    G_N &=& \frac{\kappa_5^4 \lambda}{48\pi}\nonumber\\
    \pi_{\mu\nu} &=& -\frac{1}{4} \tau_{\mu\alpha} \tau^{\alpha}{}_{\nu} + \frac{1}{12} \tau \, \tau_{\mu\nu}+ \frac{1}{8} q_{\mu\nu} \, \tau_{\alpha\beta} \tau^{\alpha\beta}- \frac{1}{24} q_{\mu\nu} \, \tau^2\nonumber
\end{eqnarray}
We haven't explicitly used the Einstein's constant in the second equation to by-pass the problem of how the brane tension relates to the Newton's constant prescription-independently. To be noted is the fact that $E_{\mu\nu}$ is the limiting case of the $y\longrightarrow0$, not the exact value of the electric part of the Weyl tensor on the brane. 
\\We'll only consider the effective brane field equation such that, there is no EM-tensor perfectly localized on the domain wall itself. However, the correspondence between the Weyl tensor and an effective EM-tensor in ordinary 4D (like tidal Reissner-Nordstrom solution) is enough to give rise to various BBH solutions. As such, we will only consider vacuum solutions of (52), thus, $T_{\mu\nu} = S_{\mu\nu} = 0$. We can also fine-tune the bulk cosmological constant such that, $\Lambda_4 = 0$. Thus, the effective brane-field Einstein equation in the limiting case becomes,
\begin{equation}
    G^\mu_\nu = -E^\mu_\nu
\end{equation}
The conservation equation of $T_{\mu\nu}$ trivially holds, on the brane. The conservation of $E_{\mu\nu}$ is given by the equation,
\begin{equation}
    D^\mu T_{\mu\nu} = 0; \text{ }D^\mu E_{\mu\nu} = \frac{48G_4\pi}{\lambda}D^\mu \pi_{\mu\nu}\nonumber
\end{equation}
Thus, in general bulk KK-models can be sourced by the variation of matter on the brane. Here, it will be satisfied as we're considering the brane is vacuum. 
\\With respect to a brane-bound observer, we can decompose the $E_{\mu\nu}$ with respect to the 4-velocity, such that,
\begin{equation}
    E_{\mu\nu} = -\left( \mathcal{U} \, u_{\mu} u_{\nu} + \frac{1}{3} \mathcal{U} \, h_{\mu\nu} + 2 q_{(\mu} u_{\nu)} + \mathcal{P}_{\mu\nu} \right)
\end{equation}
where, we have, in Planck units,
\begin{eqnarray}
\mathcal{U} &=& -E_{\mu\nu} u^{\mu} u^{\nu} \label{eq:weyl_energy} \nonumber\\
q_{\mu} &=& h^{\alpha}{}_{\mu} \, E_{\alpha\beta} \, u^{\beta} \label{eq:weyl_flux} \nonumber\\
\mathcal{P}_{\mu\nu} &=& \left( \frac{1}{3} h_{\mu\nu} h^{\alpha\beta} - h^{\alpha}{}_{(\mu} h^{\beta}{}_{\nu)} \right) E_{\alpha\beta} \label{eq:weyl_anisotropic}\nonumber\\
h_{\mu\nu} &=& g_{\mu\nu} + u_\mu u_\nu\nonumber
\end{eqnarray}
The BBH black holes will be a solution to equation the vacuum effective brane-field equation. General metric which will satisfy the equation is given by,
\begin{equation}
    ds^2 = -A(r)dt^2 + B(r)dr^2 + C^2(r)d\Omega_2^2\nonumber
\end{equation}
In general, Birkhoff's theorem don't apply to this metric - because of the fact that static solutions are not unique to gravitational collapse on the brane. As per Whisker, we are going to basically assume that there exists a timelike Killing vector of the 5D bulk metric, and we are going to work on a static slice generated by the Killing vector. Also, $C=r$ - the potential choice for area gauge - is only satisfied for all the black hole exterior if and only if $\mathcal{A} = 4\pi C^2$ increases monotonically. We have, from evaluation of the Einstein tensor components, that,
\begin{equation}
    \frac{C''}{C} = \frac{B^2}{2} \left( G^t_t - G_r^r \right) + \frac{C'}{C} \left( \frac{B'}{B} + \frac{A'}{A} \right)\nonumber
\end{equation}
This will only hold true if the DEC holds, which is a more restrictive condition than the WEC which holds true for the proof of our theorem. Since the electric part of the Weyl tensor purely generates from the gravitational effects on the bulk, it may not satisfy any energy conditions at all. Whisker \cite{Whisker2008} argues that the main reason for not choosing the area gauge is the potential presence of a wormhole region outside and around the BBH horizon, thus rendering it into a turning point into an extra reflective condition uncharacteristic of the black holes in ordinary GR (much like the horizonless condition of the Morris-Thorne wormhole). However, since we are only concerned with the quality of bound that hair may have in such braneworld scenarios, we can comfortably choose the area gauge here - since it won't change the underlying physics of the braneworld black holes.  
\\Now on to the no-short hair theorem for BBH. The ansatz metric remains the same as (1), as it is supposed to cover a wide DOC on the brane induced metric as well which has a central singularity and an event horizon. In a static spherically spacetime, $q_\mu = 0$, and $\mathcal{P}_{\mu\nu} = P(r)(r_\mu r_\nu - \frac{1}{3}h_{\mu\nu})$, where, $r_\mu$ is a unit radial vector. 
\\The corresponding Einstein field equations on the brane will be, 
\begin{eqnarray}
    \mu(\frac{1}{r^2} + \frac{1}{r}.\frac{\mu'}{\mu}) - \frac{1}{r^2} &=& - 24\pi \bar{U}(r)\nonumber\\
    \implies \mu' &=& 8\pi T^t_r r + \frac{1-\mu}{r};\text{ }T^t_t = -3\bar{U}(r)\\
    \mu(8\pi rT^t_t + \frac{1}{\mu r} - \frac{1}{r} - 2\delta' + \frac{1}{r^2}) - \frac{1}{r^2} &=& 8\pi(\bar{U}+2\bar{P})\nonumber\\
    \implies \delta' &=& \frac{4\pi r}{\mu}(T^t_t - T^r_r);\text{ }T^r_r = -\frac{T^t_t}{3} + 2\bar{P}
\end{eqnarray}
where, we have, in non-Planckian units, $\bar{U}(r) = \frac{2G_4}{(8\pi G_4)^2\lambda}\mathcal{U};\text{ }\mathcal{U} = -\frac{G_4^2}{G_5^2}E_{\mu\nu}u^\mu u^\nu$ and $\bar{P}(r) = \frac{2G_4}{(8\pi G_4)^2\lambda}P$, where, P generates from the spatially trace-free and symmetric part of $E_{\mu\nu}$, as mentioned in the decomposition of $\mathcal{P}_{\mu\nu}$. The conservation equation of the Weyl tensor will imply,
\begin{equation}
    (\mathcal{U} + 2P)' + 2 \frac{A'}{A} (2\mathcal{U} + P) = 0
\end{equation}
which is basically the same equation as this field equations. Note that, $E_{\mu\nu}$ is a trace-free tensor. This is readily seen from the general EM-tensor in terms of these quantities is given by, 
\begin{equation}
    T^\mu_\nu = (-3\bar{U}, \bar{U} + 2\bar{P}, \bar{U} - \bar{P}, \bar{U} - \bar{P});\text{ }T = 0
\end{equation}
Thus, from the field equations, and the consideration that $T = 0$, we get the modified hair function,
\begin{equation}
    e^{\delta}(e^{-\delta}r^4T^r_r)' = \frac{r^3}{2\mu}[(3\mu - 1)(T^r_r - T^t_t)]
\end{equation}
The conditions on the no-short hair theorem, and the exact statement remains the same for the BBH solutions. $T^t_t$ should fall off at $r^4$, to make sure the hair function behaves well at the asymptote. Of course, the brane metric is assumed to be asymptotically Minkowski metric. 
\subsection{The linearised metric of BBH limit}
If a point mass M is thought to be stranded on the brane, we have a linearized weak field metric on the same - black hole-like up to the first order of M - given by,
\begin{equation}
ds^2 = -\left(1 - \frac{2M}{r} - \frac{4M \ell^2}{3r^3} \right) dt^2+ \left(1 - \frac{2M}{r} - \frac{2M \ell^2}{r^3} \right)^{-1} dr^2+ r^2 d\Omega^2
\end{equation}
It can easily be seen that due to the vanishing trace property of the electric part of the Weyl tensor we are dealing with, the Ricci scalar must vanish of this solution. Indeed, calculating the Ricci scalar of the metric will yield 0 only if it is expanded upto $O(M)$. Thus, the no-short hair theorem will only be valid for the region of the BBH exterior where this solution holds, and it clearly doesn't for the entire outside-horizon area upto the asymptote. Upto $O(M)$ we have, the component of the Weyl tensors given by,
\begin{equation}
    E^t_t = -\frac{4Ml^2}{r^5};\text{ }E^r_r = -\frac{2Ml^2}{r^5};\text{ }E^r_r - E^t_t = \frac{2Ml^2}{r^5}
\end{equation}
So upto the area outside the BBH solution where this holds, the WEC holds, and so does the no-short hair theorem. However, as a check we see that,
\begin{eqnarray}
    \mu &=& \frac{1}{3} \implies \frac{2M}{r_{hair}} + \frac{2Ml^2}{r_{hair}^3} = \frac{2}{3}\nonumber\\
    r_H^3 - 2Mr_H^2 - 2Ml^2 &=& 0\nonumber\\
    r_{ph}^3 - 3Mr_{ph}^2 - \frac{10}{3}Ml^2 &=& 0\nonumber
\end{eqnarray}
The exact solutions of these equations can be found by Cardano's formula, but that will have complicated expression. For a numerical check, if we put $M=1$ and $l=0.1$, we will have, using SymPy,
\begin{eqnarray}
    r_{hair} \approx 3.0033\nonumber\\
    r_H \approx 2.0050\nonumber\\
    r_{ph} \approx 3.0037\nonumber
\end{eqnarray}
These are the only positive solutions of these cubic equations, the other two being imaginary and complementary to each other. We can see that $r_{hair}$ is slightly more in magnitude than the $r_{ph}$, implying that the approximation orders and techniques used in deriving this linearised metric is completely perturbative - and hence, the no-short hair theorem will be an observable effect iff this perturbative analysis holds for some specific values of M and l. 
\subsection{Tidal Reissner-Nordstrom Black Holes on the Brane}
Under the identification that $-E^\mu_\nu \longrightarrow T^\mu_\nu)_{EM}$ \cite{Dadhich_2000}, where the latter belongs to the ordinary RN-solution, under static spherically symmetric assumption. Thus, this identification leads to the vacuum braneworld solutions, with the identification, $g_{tt} = g_{rr}^{-1}$, 
\begin{equation}
    ds^2 = -\left(1 - \frac{2M}{r} + \frac{Q}{r^2} \right) dt^2
+ \left(1 - \frac{2M}{r} + \frac{Q}{r^2} \right)^{-1} dr^2
+ r^2 d\Omega^2
\end{equation}
Unlike classical GR, $Q < 0$ is an allowed solution here, being a tidal charge parameter, arising from the bulk gravitational effects. When, $Q>0$, it behaves as the RN-solution, there are two horizons, $r_{\pm} = M\pm \sqrt{M^2 - Q}$, if $Q < M^2$, and both of them lie within the Schwarzschild horizon, and the central singularity is timelike. However, if $Q<0$, there is only one horizon and the central singularity is spacelike - $r_H = M + \sqrt{M^2 + |Q|}$. This is the more physical solution, as this strengthens the gravitational field on the brane due to bulk effects. 
\\The Weyl tensor compoents turn out to be,
\begin{equation}
    E^t_t = \frac{Q}{r^4};\text{ }E^r_r = \frac{Q}{r^4};\text{ }E^r_r - E^t_t = 0
\end{equation}
The WEC holds always, so the no-short hair theorem holds apparently. However, the $r^4T^t_t\longrightarrow 0$ doesn't hold, it approaches $|Q|$ asymptotically. However, this can also be seen that this does not generalize to the weak field metric (60), so again, this metric cannot be the most general metric describing a BBH exterior. So, upto the range where this metric holds (way inside the asymptote) we will see, that this no-short hair theorem holds true for all the conditions as mentioned in the statement. It is seen in \cite{GuedensClancyLiddle2002} that primordial small BBHs will have this metric always due to lesser contribution from the AdS curvature, and thus - we might be able to view the short hair as phenomenlogical consequences. 
\subsection{Some Other BBH solutions}
The black string solution \cite{PhysRevD.61.065007} has the metric, by replacing the Minkowski metric in the canonical form of the RS-model metric with the Schwarzschild metric on the brane,
\begin{equation}
    ds^2 = e^{-2|y|/\ell} \left[ 
    -\left(1 - \frac{2M}{r} \right) dt^2
    + \left(1 - \frac{2M}{r} \right)^{-1} dr^2
    + r^2 d\Omega^2
\right] + dy^2
\end{equation}
This solution has $E^\mu_\nu = 0$, so the no-short hair theorem trivially holds, as in AdS-Schwarzschild black holes. However, this solution is not stable under Gregory-Laflamme conditions, but this has some interesting extra-dimension observational signatures appearing in GW reflectivity. So no-short hair theorem might be able to influence the QNM analysis of the full non-perturbative analysis of this 5D solution \cite{ClarksonSeahra2006}.
\\If we assume, $g_{tt}\text{ or, }g_{rr} = 1 - \frac{2M}{r}$, then we have two solutions due to \cite{CasadioFabbriMazzacurati2002},
\begin{equation}
    ds^2 = -\left[ \left(1 + \epsilon\right) \frac{1}{r} \left(1 - \frac{2M}{r}\right) - \epsilon \right]^2 dt^2 
+ \left(1 - \frac{2M}{r} \right)^{-1} dr^2 
+ r^2 d\Omega^2
\end{equation}
and, \cite{GermaniMaartens2001}, 
\begin{equation}
    ds^2 = -\left(1 - \frac{2M}{r} \right) dt^2
+ \frac{\left(1 - \frac{3M}{2r} \right)} {\left(1 - \frac{2M}{r} \right) \left(1 - \frac{r_0}{r} \right)} dr^2
+ r^2 d\Omega^2
\end{equation}
Both of these solutions will perturbatively satisfy the no-short hair theorem away from the far-field limit - but since these solutions have no unique photon sphere radius, we won't be able to exactly state that whether the no-short hair limit will lie outside the photon sphere radius. 
\\Technically speaking, from the metric (65), we get, $G^r_r - G^t_t =
\frac{2[(1+\epsilon)(6M^2 + r^2) - Mr(\epsilon r + 5\epsilon + 5)]}{r^3[2M(1+\epsilon) + \epsilon r^2 - \epsilon r - r]}$. Since clearly this has a negative limit for $\epsilon = 0$, which is given by, $-\frac{1}{4M^2}$, there can be no possible real $\epsilon$ which will satisfy the WEC everywhere outside the black hole horizon in the far-field limit. 
\\In a similar manner, we can never find a value of $r_0$ which will allow the WEC satisfied in all of the far-field limit of the braneworld black hole. 
\section{Braneworld Wormholes}
The black hole solutions we worked in previously in the braneworld scenario, are all prevalently done in the RS II model - as all of the inherent geometries of the solution imply that we live on the negative tension brane in RS I model and therefore that violates our previously known observations of the world in terms of the particle zoo. However, there is an effective scalar-tensor theory constructed by \cite{KannoSoda2002,ShiromizuKoyama2003} on the braneworld that we get in the RS I model. The background is - the two 3-branes are located at $y=0$ and $y=l$, where y denotes the extra-dimension warped. Since this effective theory is no longer valid for extreme cases like black holes, whatever generalizations done in this section will not translate to the previous sections. But since wormhole entails finite energy densities and pressures, this theory holds and the effective field equations is given by,
\begin{eqnarray}
    G_{\mu\nu} &=&
\frac{\bar{\kappa}^2}{\,l\Phi}\,T^{b}_{\mu\nu}
+ \frac{\bar{\kappa}^2}{\,(1+\Phi)\,l\Phi}\,T^{a}_{\mu\nu}
+ \frac{1}{\Phi}\left( \nabla_{\mu}\nabla_{\nu}\Phi - g_{\mu\nu}\nabla_{\alpha}\nabla^{\alpha}\Phi \right)\nonumber\\
&-& \frac{3}{2\,\Phi(1+\Phi)}\left(
\nabla_{\mu}\Phi\,\nabla_{\nu}\Phi
- \frac{1}{2}g_{\mu\nu}\nabla_{\alpha}\Phi\,\nabla^{\alpha}\Phi
\right)
\end{eqnarray}
To make the theory truly scalar-tensor, we assume that the EM-tensor on the Planck brane ($T^a_{\mu\nu}$) is identically zero. 
\\The radion field is a measure of the distance between the two 3-branes. If we denote the four-vectors on the brane as x, we have, $\Phi(x) = e^{\frac{2d(x)}{l}} - 1$, where, d is the proper distance between the two 3-branes given as, 
\begin{equation}
    d(x) = \int^l_0 e^{\phi(x)}dy\nonumber
\end{equation}
with $\phi$ appearing in the usual 5D line element of the brane AdS space, the warping factor. The scalar radion field satisfies the equation,
\begin{equation}
    \nabla_{\alpha}\nabla^{\alpha} \Phi =
\frac{\bar{\kappa}^2}{\,l}\frac{T^{(a)} + T^{(b)}}{2\omega + 3}
- \frac{1}{2\omega + 3}\,\frac{d\omega}{d\Phi}
\left( \nabla_{\alpha}\Phi \right)\left( \nabla^{\alpha}\Phi \right)
\end{equation}
where, $\omega(\phi) = -\frac{3\Phi}{2(1+\Phi)}$. 
\\Both of these equations is exactly in the same form of scalar-tensor theory as discussed before. Therefore, that limits the number of theories of scalar-tensor class which relates the conservation of EM-tensor on the visible brane as before, but as long as that is satisfied, we are good to go. We assume that $\Phi$ is never zero, and $\Phi$ does not diverge to infinity at any finite value of the brane coordinate. These will ensure that we have a stable radion.
\\The field equation can be effectively written as,
\begin{equation}
    G_{\mu\nu} =
\frac{\bar{\kappa}^2}{l\Phi}\,T^{b}_{\mu\nu}
+ \frac{1}{\Phi}\,T^{\Phi}_{\mu\nu}
\end{equation}
The Raychaudhuri equation for timelike expansion coefficient $\Theta$ for congruences is given by,
\begin{equation}
    \frac{d\Theta}{d\lambda} + \frac{1}{3}\Theta^2 + \Sigma^2 - \Omega^2 = -R_{\mu\nu}u^\mu u^\nu
\end{equation}
where, $\Omega_{ij}$ is the rotation, and $\Sigma_{ij}$ is the shear, and $\lambda$ is the affine parameter. We know that for any matter present on the spacetime, we will have the timelike geodesic focus around that to some extent, making the RHS is always greater than or equal to zero. In GR, this congruence implies the SEC, and different versions of the same congruence leads to WEC and NEC which lead to physical requirements such as the energy measured in every reference frame being strictly positive. We put the effective field equation (69) in the Raychaudhuri equation to see that,
\begin{equation}
    R_{\mu\nu}u^{\mu}u^{\nu} =
\frac{\bar{\kappa}^2}{l\Phi}\,T^{b}_{\mu\nu}u^{\mu}u^{\nu}
+ \frac{1}{\Phi}\,T^{\Phi}_{\mu\nu}u^{\mu}u^{\nu}
\geq 0\nonumber
\end{equation}
Thus, it is mathematically possible to satisfy the congruence condition and still violate WEC for the brane EM-tensor and vice-versa, due to the effective stress-energy tensor appearing due to the radion field. Like the Newton's constant variation being practically unobservable to the Jordan frame observer due to the fact this effects occur at cosmic scales and virtually impactless at ordinary observers, the radion field stress-energy tensor has also got nothing to do with the matter content on the brane.
\\The specific wormhole solution considered here will be characterized by the 4-Ricci scalar being zero, thus, the brane EM-tensor being traceless. So, the satisfaction of the no-short hair theorem will completely depend on the WEC being true. So, we have to find a solution for which there is no on-brane exotic matter content on the wormhole.
\\In a general form, in isotropic coordinates, a static spherically symmetric geometry is given by (see Appendix C for a brief discussion),
\begin{equation}
    ds^{2} = -\frac{f^{2}(r)}{U^{2}(r)}\,dt^{2}
+ U^{2}(r)[\,dr^{2}
+ r^{2}\,d\theta^{2}
+ r^{2}\sin^{2}\theta\,d\phi^{2}]
\end{equation}
Using the field equations, we will get a solution of a wormhole on the brane and using the traceless condition (the radion field will have no contribution from the on-brane matter because $R =0$, for $T^b_{\mu\nu}$, thus satisfying our condition for $\Box\phi$ in BD-form gravity), 
\begin{eqnarray}
    f(r) &=& \left( 1 + \frac{m}{2r} \right)
\left[ \kappa( 1 + \frac{m}{2r}) + \lambda \left( 1 - \frac{m}{2r} \right) \right]\nonumber\\
U(r) &=& \left( 1 + \frac{m}{2r} \right)^{2}
\end{eqnarray}
where, $\kappa > \lambda > 0$ and the wormhole throat is present at $r' = 2m$, and where, $r' = r(1+\frac{m}{2r})^2$, the former being the Schwarzschild coordinate. 
\\We can find $\xi = \sqrt{1+\Phi}$ from f and the radion field equation, to get, 
\begin{equation}
    \xi = \frac{C_{1}}{m\lambda} \ln\!\left( \frac{2r q + m}{2r + m} \right) + C_4
\end{equation}
where, $q = \frac{\kappa + \lambda}{\kappa - \lambda}$. To have a well-behaved radion, we need to make sure that $\Phi$ is never zero or doesn't blow up anywhere for any value of r. To make sure of that, we choose, $C_1, C_4 > 0\text{, and }q>1$. 
\\where, $x = \frac{m}{2r}\text{, }C_1 = \alpha m\text{, }\beta = \frac{\alpha}{\kappa - \lambda}$. The domain of x is from 0 to 1, as the wormhole throat is at $r = \frac{m}{2}$. 
\\The radion field can be found out to be given by,
\begin{equation}
    \xi(x) = \sqrt{1 + \Phi} =
\frac{\alpha}{\lambda}
\ln\!\left( \frac{q + x}{1 + x} \right) + C_4
\end{equation}
where, $x = \frac{m}{2r}\text{ and }C_1 = \alpha m\text{, }\beta = \frac{\alpha}{\kappa - \lambda}$ are parameters that appear in the field equations of $\rho$ and $p_r$. 
This outline is detailed in \cite{KarLahiriSenGupta2015}. Plotting graphs for the energy condition expressions in terms x with respect to the evolution of the parameters $q, \beta, C_4$ we can choose the values of these parameters such that the WEC will be satisfied by the on-brane matter. To make the term inside the square bracket go to zero (to satisfy the more general WEC being satisfied for the tangential pressure as well), we have to make sure that $\beta$ and $q$ satisfy an inter-relation given by,
\begin{equation}
    \beta^{2} =
\frac{(q + 1)(q - 1)^2}
{4\left[ (q + 1)(\ln q)^{2} - 2(q - 1)\ln q \right]}\nonumber
\end{equation}
With the choice of $m = 1, q = 3, C_4 = 0$, we can uniquely determine $\beta$ and graphically show that this satisfies the WEC both far and away from $r = \frac{m}{2}$, that is covering the entire space. Also, the choice keeps the radion field well-behaved through all the possible range of r. Any other choice of $C_4 > 0, \beta > 0, q>1$ may violate the WEC over a wide range of x. 
\\Thus, the no-short hair theorem is trivially satisfied in this example of a braneworld wormhole. The only thing left to check if, upon identification of the condition on $\omega(\phi)$\footnote{Upon analogy to 4D Brans-Dicke theory of gravity, done in Appendix A.} is, $\omega' = \frac{d\omega}{d\Phi} < 0$ is satisfied - which, we can easily check that, $\omega' = -\frac{6}{4(1+\Phi)^2}$ which is always less than zero as long as the radion field is well behaved. 
\section{Discussion}
To summarize the conditions of no-short hair theorem to be valid in theories of Post-Einstein gravity:-
\begin{itemize}
    \item 5D EGB theory will have black hole solutions satisfying the no-short hair theorem, iff, $\alpha > 0$.
    \item 4D (and higher-dimensional) Brans-Dicke theory will allow BH solutions which satisfy the no-short hair theorem if for $\omega$ constant, we have $\Box \phi \geq 0$ and if $\omega = \omega(\phi)$ then we have, in addition to the previous condition, $\omega' = \frac{d\omega}{d\phi} < 0$. 
    \item Tidal Reissner-Nordstrom BH solution for braneworld scenario in RS-2 model will not allow no-short hair behavior. However, some other braneworld black holes like black string and linearised BBH do allow no-short hair behavior under very stringent conditions; however, the WEC and other conditions that result into the no-short hair theorem may not hold for all the far-field limit.  
    \item Galactic isolated wormholes (with dark matter halo, galactic wormholes may show no-short hair behavior - see \cite{Biswas_2024} for a discussion on energy condition and light-ring positions for galactic DS and braneworld wormholes; this may be taken up in a future work to explicitly see if this no-short hair theorem generalizes to these cases) and traversable wormholes will not show no-short hair behavior; however, braneworld wormholes in RS-1 scenario will show no-short hair behavior, owing to the WEC-satisfaction of the on-brane observer in the Jordan frame within Kanno-Soda effective gravity theory in RS1 braneworld. This can be a probe for a robust observational proof of extra-dimensions if wormholes are ever observed in our universe. 
\end{itemize}
As is well known, if $\alpha > 0$ in EGB, then it is the inverse string tension related to the low energy limit of the higher-dimensional string theory. Thus, the no-short hair theorem is generalizable in heterotic string theory as well, presumably. 
\\In fact, this $\alpha>0$ condition might be a more subtle condition for the Gauss-Bonnet theory's practicality. The cosmological action just adds up a linear term with the cosmological constant in the hair function, so without losing or adding any physics related to the no-short hair theorem, we can consider the more general action,
\begin{equation}
    S = \frac{1}{16\pi G}\int d^5x\sqrt{-g}(R - 2\Lambda + \alpha L_{GB}) + S_{matter}
\end{equation}
We expand the metric about a maximally symmetric spacetime ($\bar{g}_{\mu\nu}$) with radius of curvature $l$, and $k = 1/l^2$, so we get,
\begin{eqnarray}
    g_{\mu\nu} = \bar{g}_{\mu\nu} + h_{\mu\nu}\nonumber\\
    \bar{R}_{\mu\nu\rho\sigma} = k \left( 
    \bar{g}_{\mu\rho} \bar{g}_{\nu\sigma} 
    - 
    \bar{g}_{\mu\sigma} \bar{g}_{\nu\rho} 
\right)\nonumber\\
\bar{R}_{\mu\nu} = 4k \, \bar{g}_{\mu\nu}\nonumber\\
\bar{R} = 20k\nonumber
\end{eqnarray}
We will restrict ourselves to the TT-gauge, $\nabla^\mu h_{\mu\nu} = 0$ and $h^\mu_\mu = 0$. We will check the presence of Ostrogradsky ghost by checking the effective second order kinetic term of these metric perturbation.
\\Using standard linearized Ricci tensor and Riemann tensor techniques (as in gravitational waves) and the symmetry of the perturbed metric fluctuation tensor, we find that,
\begin{equation}
    \delta R_{\mu\nu}|_{TT} = -\frac{1}{2}(\bar{\Box}h_{\mu\nu} + 2kh_{\mu\nu})
\end{equation}
The Ricci scalar variation vanishes identically. 
\\If we Taylor expand the Einstein action (or, the EGB action) the first order vanishes because of the traceless gauge condition. In the second order, we get,
\begin{eqnarray}
    S^{(2)}_{\text{EH, TT}} 
= \frac{1}{16\pi G} 
\int d^{5}x \, \sqrt{-\bar{g}} \,
h_{\mu\nu} \left( -\bar{\Box} - 2k \right) h^{\mu\nu}\nonumber\\
S^{(2)}_{\text{EGB, TT}} = \frac{\alpha}{48\pi G} 
\int d^{5}x \, \sqrt{-\bar{g}} \,
h_{\mu\nu} \left( -\bar{\Box} - 2k \right) h^{\mu\nu}
\end{eqnarray}
Thus, the total variation in the second order of the action results in the usual Lichnerowicz operator, with the Gauss-Bonnet coefficient only affecting the prefactor. The prefactor (say $\kappa$) turns out to be,
\begin{equation}
    \kappa = 1 + \frac{\alpha k}{3} = 1+\frac{\alpha}{3l^2}
\end{equation}
If the background geometry is flat, then $\alpha$'s sign is irrelevant - however, if we consider EGB gravity in a cosmological background (which is our real world) the positive sign of $\alpha$ makes sure that the kinetic prefactor is always positive, getting rid of the ghost propagators (characteristic of the Gauss-Bonnet term unlike other higher curvature terms which result in extra-mode propagators at higher dimensions). Thus, if the Gauss-Bonnet gravity model we are considering has no ghost propagators, the BH solutions it allows will have no-short hair behavior. A more complete proof involving a total UV-complete EFT description of EGB including massive and tachyonic spectral modes for $\alpha>0$ can be found at \cite{Cheung_2017}. 
\\It can also easily be seen from the 6D EGB calculation, that all that changes in the hair function is the magnitude of the prefactors of the terms in the RHS of the expression, where the steps are very analogous to the 5D one, thus $\alpha>0$ being the only condition for the satisfaction of no-short hair theorem as well. And since, only 5D and 6D are the relevant cases in the low energy limit (considering that only these two dimensions unfurl before the theory reverts back to higher-order supergravity corrections at higher energies in 7D) - we can conjecture that, \textit{all fully quantum black holes must obey the no-short hair theorem for quantum hairs in quantum theories of gravity}.
\\Even though in Appendix B, we have been able to calculate the hair-bound in a theory-agnostic manner and find its relation to null circular orbits, this procedure has several conceptual problems. Firstly, this analysis will not hold for asymptotically non-flat cases, as the behavior at infinity is assumed to be flat and that supplies part of the reason where this photon sphere radius becomes the hair-bound. This procedure ad hoc assumes non-minimal coupling; however, in BD theory derivation in Appendix A, we have seen that non-minimal coupling may lead to additional constraints (albeit simple) on the matter field EM-tensor. Thirdly, horizons in general are not very well-behaved - that is, can be fuzzy or effective - and hence causal effects may not be very easy to compute in a semiclassical or quantum theory of gravity in an theory-agnostic fashion. In general, the Raychaudhuri equation assumed for null congruence (letting hairy effects on geodesics to be focused out of light ring as well) to be true, assumes monotonicity of the metric functions and absence of exotic defocusing effects - higher-derivative theories inspired by quantum gravity sometimes may result into violation of Raychaudhuri equation\cite{doi:10.1142/S0219887821501152, Ahmadi_2006}, or even cosmological scenarios. A comprehensive study of no-hair theorems for cosmological BHs has been done in \cite{ishibashi2024notenohairpropertiesstatic}, in complement to \cite{AuthorMastersThesis}. In particular, one of the central results of the latter was the claim of BH hair continuing till cosmological horizon under the assumption of the dark matter content being bound on the lower side by one-fourth of the trace of matter field EM-tensor of the universe in dS space is supposed to change for modified gravity in dS space. Some of these results might be taken up by the author in future works. Thus, the approach by \cite{Hod_2011} is more suitable, in complement with the formalism developed by this present work, because explicit use of equations of motion in modified gravity theories lets us have more control over the parameters of the gravitational theory and understand why and how the theorem affects black hole solutions in those theories. 
\\In general, in semiclassical and quantum theories of gravity, the fundamental energy condition - the NEC - is replaced by a quantum version, called QNEC, given by, $\langle T_{kk} \rangle\;\ge\;\frac{1}{2\pi}\,\frac{d^{2} S_{\text{ent}}}{d\lambda^{2}}$, where, $S_{ent}$ is the entanglement entropy of quantum fields across a cut of the null hypersurface - owing to Hawking radiation, Casimir effect, and so on\cite{iizuka2025energyconditionsquantuminformation}. In other words, energy density can be negative if suitably compensated by quantum informational structure of the spacetime. This provides the inequality bounds for QWEC, where we have the quantum focusing conjecture to be true and that may lead to negative energy measured by a TL observer within a certain bound. In general, in those scenarios, if a consistent quantum no-short hair theorem can be found out to be true, then - considering that the entropy of a black hole is directly proportional to the amount of hair it has - the holographic principle bound may be violated, that is, $S_{max} = \frac{A}{4G_N}$\cite{Ryu_2006}. This direction of inquiry will be taken up by the author in some future work. 
\\It is an interesting direction to observe at the subset of $f(R)$ theories of gravity under scalar-tensor (taking inspiration from Appendix A) to understand the no-short hair behavior in higher derivative theories of gravity and also under conformal transformation - those results will be shown in a future paper by the author. 
\\In principle, one can observationally verify the obeyance of the no-short hair theorem at suitable energy scales by local quantities defined well outside supposedly hairy black holes and seeing if they require other charges to be described, through LIGO-VIRGO collaboration. 
\section{Acknowledgment}
The author would like to acknowledge the guidance of and the materials provided by Profs. Soumitra SenGupta and Sumanta Chakraborty during the pursuance of this article - this article covers part of the Master's research done by the author for his Integrated BS-MS course under the aforementioned professors' supervision. The author would gratefully acknowledge discussions with Prof. Sourov Roy, and his IACS colleagues for being such a wonderful host. The author would acknowledge Rosa Laura-Lechuga Solis for a calculational clarification. The author was funded by a fellowship granted for Master's Student under IACS from DST, MHRD, Government of India throughout the duration of this research. 
\section{Appendix A - The No-Short Hair Theorem in Brans-Dicke Gravity}
Brans-Dicke action \cite{BransDicke1961} was proposed as an alternative to general relativity based on Mach's principle of absolute relativity. It replaces the Newton's constant by a scalar coupling with the Ricci curvature term non-minimally, much akin to the dilatonic field. The action is given by,
\begin{equation}
    S = \frac{1}{16\pi}\sqrt{-g}(\phi R - \frac{\omega}{\phi}\partial_\mu\phi\partial^\mu\phi) + S_M
\end{equation}
$\omega$ is a constant (by experimental signatures, $|\omega| > 40,000$), and $\phi$ is a scalar field is generated by the trace of the stress-energy tensor. In the case of static spherically symmetric metric of a black hole, we will have $\phi$ will only be radial. Variation of the action with respect to the metric tensor gives us,
\begin{equation}
    G_{\mu\nu} = \frac{8\pi}{\phi}T_{\mu\nu} +\frac{\omega}{\phi^2}(\partial_\mu\phi\partial_\nu\phi - \frac{1}{2}g_{\mu\nu}\partial_\lambda\phi\partial^\lambda\phi) + \frac{1}{\phi}(\nabla_\mu\nabla_\nu\phi - g_{\mu\nu}\Box\phi)
\end{equation}
Taking the trace of equation (39), and varying the action with respect to the scalar field, and adding both the contributions together, we get,
\begin{equation}
    \Box\phi = \frac{8\pi}{3+2\omega}T
\end{equation}
We will have the following equations by easy investigation,
\begin{eqnarray}
    \partial_\lambda\phi\partial^\lambda\phi = \mu\phi'^2\\
    \nabla^r\nabla_r\phi = \mu(\phi'' + \frac{\mu'}{2\mu}\phi') = \frac{8\pi T}{3+2\omega}\\
\end{eqnarray}
In general, we will consider that $\frac{\delta T^m_{\mu\nu}}{\delta\phi} = 0$. Thus, diffeomorphism invariance will imply, $\nabla_\mu T^{\mu\nu}_m = 0$. Then, to make sure that the effective conservation equation holds, $\nabla^\mu T_{\mu\nu}^\phi = \frac{8\pi}{\phi^2}T^m_{\mu\nu}\partial^\mu\phi$ - meaning the EM-tensor of scalar field in the presence of matter fields will not satisfy a conservation equation by itself. However, here we work with the former, as on-brane matter has no contribution whatsoever in the radion in RS1 model because of the tracelessness of the EM-tensor of the matter fields on the brane. If we want to work with the full non-minimal coupling, all that will be required as an extra-assumption, both now and in the scalar-tensor form is an extra condition that (which can be straightforwardly seen), $P_r)_m < 0$.
\\After getting the Einstein-Brans-Dicke equations here, we will have, analogous to the 4D GR calculation, the hair function turning out to be,
\begin{eqnarray}
    e^\delta(e^{-\delta}r^4T^r_r)' &=& \frac{r^3}{2\mu}[(3\mu - 1)(T^r_r - T^t_t) + 2\mu T] + (T^r_r - T^t_t)(\frac{1}{4}\frac{\omega}{\phi^2}\phi'^2 \nonumber\\
    &+& \frac{4\pi T}{(3+2\omega)\mu\phi})r^5 + T^t_t r^5[\frac{\omega}{2\phi^2}\phi'^2 + \frac{4\pi T}{(3+2\omega)\mu\phi}]
\end{eqnarray}
\\The last additive term on the RHS of the equation is always negative, as long as, $\phi' < 0$, and WEC holds on the effective EM-tensor as well. We will resort to a lesser parsimonious representation of the field equation. We have, by definition,
\begin{equation}
    G^\mu_\nu = \frac{8\pi}{\phi}T^\mu_\nu)_m + \frac{1}{\phi}T^\mu_\nu)_\phi
\end{equation}
That results into saying that,
\begin{equation}
    T_{\mu\nu})_\phi = \frac{\omega}{\phi}(\partial_\mu\phi \partial_\nu\phi - \frac{1}{2}g_{\mu\nu}\partial^\alpha\phi\partial_\alpha\phi) + \nabla_\mu\nabla_\nu\phi - g_{\mu\nu}\Box\phi
\end{equation}
From here on, we will denote the trace of matter EM-tensor to be $T^m$ and that of the scalar field EM-tensor to be $T^\phi$. We thus have,
\begin{equation}
    T^\phi = \frac{\omega}{\phi}(-\mu\phi'^2) - \frac{24\pi T^m}{3+2\omega}
\end{equation}
Using the definition of the scalar field EM-tensor, the second and third coefficients can be rewritten as,
\begin{eqnarray}
    \frac{\omega}{4\phi^2}\phi'^2 + \frac{4\pi T^m}{(3+2\omega)\mu\phi} &=& \frac{1}{\mu\phi}(-\frac{T^\phi}{2} - \frac{8\pi T^m}{(3+2\omega)})\nonumber\\
    \frac{\omega}{2\phi^2}\phi'^2 + \frac{4\pi T^m}{(3+2\omega)\mu\phi} &=& \frac{1}{\mu\phi}(-\frac{T^\phi}{4} - \frac{2\pi T^m}{(3+2\omega)})\nonumber
\end{eqnarray}
Since we have to make sure that the RHS of equation (82) must be negative at least upto the point where $(3\mu - 1) > 0$, we will have a combination of conditions being either of these equations being greater than zero or lesser than zero, which then asserts us, that (note, $T^m < 0$, by condition), iff,
\begin{equation}
    \frac{8\pi |T^m|}{3+2\omega} < T^\phi < \frac{16\pi |T^m|}{3+2\omega}
\end{equation}
the no-short hair theorem will be satisfied. This gives us a constraint on $\omega$, which does hold if we take a positive value of $\omega$. Note that, $T^\phi$ can be positive in this case, as per equation (89), as long as $\omega$ satisfies some bound. 
\\However, there are too much unknowns here for these terms to juggle by. To get a constraint on $\phi$, or even to understand what is going on, a good proposition would be to drop the $T^m$ term, like the dilaton EM-tensor we took in the case studies, with no coupling. We can work with a convex potential term, but since it is convex, the first derivative of that potential with respect to $\phi$ will always be negative, so it won't qualitatively contribute to theorem's condition or not. So, we drop that too, here.
\\So, altogether, we have, in a modified equation of motion for vacuum, and the effective Bianchi identity,
\begin{equation}
    G^\mu_\nu = \frac{1}{\phi}T^\mu_\nu)_\phi \implies \nabla_\mu T^\mu_\nu)^{eff}_\phi = 0; \Box\phi = 0
\end{equation}
Thus, the new hair function becomes,
\begin{equation}
    e^\delta(e^{-\delta}r^4T^r_r)_\phi)' = \frac{r^3}{2\mu}[(3\mu - 1)(T^r_r)_\phi - T^t_t)_\phi) + 2\mu T^\phi]
\end{equation}
As a check, we can see that,
\begin{eqnarray}
    T^\phi = -\omega\mu\frac{\phi'^2}{\phi} &<& 0\nonumber\\
    T^r_r)_\phi - T^t_t)_\phi &=& \frac{\omega}{\phi}\mu\phi'^2 + \mu\phi'' + \frac{1}{2}\mu'\phi' = -T^\phi + \Box\phi\nonumber 
\end{eqnarray}
Thus, the no-short hair theorem will be satisfied iff,
\begin{equation}
    \mu(\phi'' + \frac{\mu'}{2\mu}\phi') = \Box\phi \implies \Box\phi \leq -T^\phi
\end{equation}
Thus, this is the constraint on the scalar field. If this is satisfied, and is translated to the $T^\mu_\nu)_m \neq 0$ case, we will have, $\Box \phi = \frac{8\pi T^m}{3+2\omega}< 0$, consistent with our original condition. 
\subsection{4D scalar-tensor theory of gravity}
A more general form of the Brans-Dicke theory of gravity is when $\omega = \omega(\phi)$. The action is, therefore, in this context,
\begin{equation}
    S = \frac{1}{16\pi}\sqrt{-g}(\phi R - \frac{\omega(\phi)}{\phi}\partial_\mu\phi\partial^\mu\phi) + S_M
\end{equation}
The equations of motion of both the metric field, and the scalar field, are,
\begin{equation}
    G^\mu_\nu = \frac{8\pi}{\phi}T^\mu_\nu + \frac{1}{\phi}[\nabla^\mu\nabla_\nu\phi - \delta^\mu_\nu\Box\phi] + \frac{\omega(\phi)}{\phi^2}(\partial^\mu\phi\partial_\nu\phi - \frac{\delta^\mu_\nu}{2}\partial^\alpha\phi\partial_\alpha\phi)
\end{equation}
\begin{equation}
    \Box\phi = 8\pi\frac{T}{(3+2\omega)} - \frac{\omega'}{(3+2\omega)}(\partial^\mu\phi\partial_\mu\phi)
\end{equation}
Thus, the equation of motions for the relevant components of the spacetime field,
\begin{eqnarray}
    r\delta' &=& \frac{4\pi r^2}{\mu\phi}(T^t_t - T^r_r) - \frac{\omega}{2\phi^2}r^2\phi'^2 + \frac{r^2}{\phi}(\frac{\omega'\phi'}{2(3+2\omega)} - \frac{4\pi T}{\mu(3+2\omega)})\nonumber\\
    \frac{r\mu'}{2\mu} &=& \frac{1}{2\mu}(\frac{8\pi}{\phi}T^t_t r^2 + \frac{1}{\phi}(\frac{\omega'r^2\mu\phi'^2}{3+2\omega} - \frac{8\pi Tr^2}{3+2\omega}) - \frac{1}{2}\frac{\omega r^2}{\phi^2}(\mu\phi'^2) + 1 - \mu)\nonumber
\end{eqnarray}
The hair function thus becomes,
\begin{eqnarray}
    e^\delta(e^{-\delta}r^4T^r_r)' &=& \frac{r^3}{2\mu}[(3\mu - 1)(T^r_r - T^t_t) + 2\mu T] + (T^r_r - T^t_t)(\frac{1}{4}\frac{\omega}{\phi^2}\phi'^2 \nonumber\\
    &+& \frac{4\pi T}{(3+2\omega)\mu\phi} - \frac{1}{\phi}\frac{\omega'}{2(3+2\omega)}\phi'^2)r^5 + T^t_t r^5[\frac{\omega}{2\phi^2}\phi'^2 + \frac{4\pi T}{(3+2\omega)\mu\phi} \nonumber\\
    &-& \frac{\omega'\phi'^2}{2(3+2\omega)\phi}]
\end{eqnarray}
where, $T^\phi$ will be given by,
\begin{equation}
    T^\phi = \frac{\omega}{\phi}(-\mu\phi'^2) - \frac{24\pi T^m}{(3+2\omega)} + \frac{3\omega'}{(3+2\omega)}\mu\phi'^2\nonumber
\end{equation}
It is easy to see that the no-short hair theorem will hold iff,
\begin{equation}
    -\frac{1}{3+2\omega}(8\pi T^m + \frac{1}{4}\omega'\mu\phi'^2) < T^\phi < -\frac{1}{3+2\omega}(16\pi T^m + \omega'\mu\phi'^2)
\end{equation}
But, again, here too many unknowns are at play and getting a consistent constraint will be difficult from this set-up. So, as before, we drop the matter EM-tensor, and redo the whole procedure to get to the point,
\begin{equation}
    \Box\phi \leq -T^\phi - \frac{\omega'}{3+2\omega}\mu\phi'^2\implies -\frac{\omega'}{3+2\omega}\mu\phi'^2 > 0\text{ (By Occam's Razor)}
\end{equation}
This implies that, $\omega' < 0$ for the no-short hair theorem to be satisfied. The conditions for the constrained conservation equation translates from the previous section as well. 
\\It will be straightforward but tedious to show that this conditions generalize to higher dimensions.
\section{Appendix B - A Heuristic Theory-Agnostic Derivation of the No-Short Hair Bound}
Null circular orbits are defined as the last stable orbit from where the gravity is so strong that even the photon takes a round trajectory there, around a black hole. \cite{Carroll2004}Now, in general, it arises a neighbourhood of the event horizon of a black hole where gravity is so strong that emitted photons will not just bend around the black hole but also return to the point where they were emitted from and consequently display boomerang-like properties. Therefore, the innermost null circular orbit is the lowest bound for any stable orbit around that black hole, conveying that timelike geodesics lose stable circular orbits as it approaches this orbit, even in terms of proper Schwarzschild distance in geodesics generated by the Killing vectors. In other words, in the area between the horizon and the innermost null circular orbit, there will be no other stable circular orbits for TL geodesics. 
\\A general metric in d-dimensions for a static spherically symmetric black holes are,
\begin{equation}
    ds^2 = -f(r)dt^2 + k(r)^{-1}dr^2 + r^2d\Omega_{d-2}
\end{equation}
Of course, $f(r_H) = 0$. We assume that there is at least one closed Killing vector orbit other than the one coming due to t-like isometry\footnote{This is not always true for $d\geq5$, even when the spherical Einsteinan factorizable manifold in the metric is topologically compact.}. 
\\Now, for photons, by definition, $ds^2$ is 0. So, we get,
\begin{equation}
    0 = -f(r)\dot{t}^2 + k(r)^{-1}\dot{r}^2 + r^2\dot{\phi}^2
\end{equation}
In the last step, I used the fact that there is a $\phi$-like isometry of the metric, where, $\phi$ is the last angular coordinate of the system in which the metric is defined, and there is a t-like isometry. So, all the temporal derivative of the angular coordinates except $\phi$, with respect to an arbitrary affine parameter is zero.
\\Now, the Lagrangian of a particle (in our case, it will be null-like) traveling in this geodesic will be given by,
\begin{equation}
    L = (u^\mu u_\mu)^{1/2} = 0 (\text{for photons})
\end{equation}
Since, L is a scalar field in spacetime in principle, its square will also be a scalar field in spacetime, thus satisfying the Euler-Lagrange equations. So, $L^2$ is the Lagrangian for us, for now. So, we have,
\begin{equation}
    L)_{new} = -f(r)\dot{t}^2 + k(r)^{-1}\dot{r}^2 + r^2\dot{\phi}^2
\end{equation}
In that Lagrangian, corresponding to the t and $\phi$-like isometries, the conserved quantities will be,
\begin{eqnarray}
    \frac{\partial L}{\partial \dot{t}} &=& -2f\dot{t} = \text{constant = E(let)}\nonumber\\
    \frac{\partial L}{\partial \dot{\phi}} &=& 2r^2\dot{\phi} = \text{constant = l(let)}\nonumber\\
\end{eqnarray}
Using these constants to rewrite $\dot{t}$ and $\dot{\phi}$ in (88), we have,
\begin{equation}
    \dot{r}^2 = E^2 - \frac{fl^2}{r^2}
\end{equation}
The LHS being the kinetic energy of the photon (effectively), and $E^2$ being a measure of the total energy of the photon as it trajects the orbit in a conserved system, we have, the second term of the RHS to be the effective potential of the system. This effective potential, therefore, must have a minima at the stable orbit around the black hole, that is, 
\begin{eqnarray}
    \frac{dV_{eff}}{dr} = 0 &=& h^2(\frac{f'}{r^2} - \frac{2f}{r^3})\nonumber\\
    \implies f'r &=& 2f
\end{eqnarray}
\subsection{The Actual Derivation}
Let us now say, that, for a static spherically symmetric black hole metric, which asymptotically converges to a Minkowski metric, with all the metric condition coefficients remaining the same as per the general relativity case, is given by, in 4D,
\begin{equation}
    ds^2 = -f(r)dt^2 + \frac{1}{k(r)}dr^2 + r^2d\Omega^2
\end{equation}
with, $k(r_H) = 0$; $f(r_H)$ is finite and regular, the EM-tensor is finite and regular at the horizon, and $f(r) = k(r) = 1$ at infinity. 
We will find an important component of the calculation,
\begin{equation}
    \frac{1}{\sqrt{-g}}\partial_r\sqrt{-g} = \frac{1}{r} + \frac{1}{2}(\frac{f'}{f} - \frac{k'}{k})
\end{equation}
Then, as usual, we demand the conservation of EM-tensor, $\nabla_\mu T^\mu_\nu = 0$, and we pour out for the only non-trivial component, we get,
\begin{equation}
    \nabla_rT^r_r = 0 = \partial_rT^r_r + (\frac{1}{\sqrt{-g}}\partial_r\sqrt{-g})T^r_r - (\Gamma^t_{tr}T^t_t + \Gamma^r_{rr}T^r_r + \Gamma^{\theta}_{\theta r}T^\theta_{\theta} + \Gamma^{\phi}_{\phi r}T^\phi_\phi)
\end{equation}
Putting the necessary components' substitution, and progressing along the same lines of computation as we did in the 4D GR case (though not substituting $f'$ and $k'$ with anything as this will come from the theory of gravity underlying), we will get a modified hair function, such that,
\begin{equation}
    (r^4T^r_r)' = \frac{r^3}{2f}[(T^r_r - T^t_t)(2f - rf') + 2fT]
\end{equation}
Thus, it is clearly visible that if the WEC is satisfied, and the trace of the matter EM-tensor is less than zero, then, the no-short hair theorem will be trivially satisfied as per the GR case, with the bound on $f$ (the $g_{tt}$ component of the metric), as, $2f - rf'$. This is nothing but the expression of the null circular orbit of the black hole if solved with an equality. If the metric allows more than one solution to the photon sphere equation, we will generally have to take the lowermost one as the hair bound. It is because, by Rolle's theorem, the LHS of equation (106) being a continuous function, at each null circular orbit the function will change its sign - and thus, effectively, the hairy behavior might be locally violated between the region of two null circular orbits (of course, after the outermost null circular orbit we see the function uniformly becoming positive). This can be argued against by demanding that if hair continues till the innermost null circular orbit, the presence of other null circular orbits should not change the physics of the hair and hence the BH solution should satisfy the no-short hair theorem outside that bound uniformly to an asymptotic observer. In other words, this means that between the region of the BH horizon and the innermost null circular orbit, the LHS of equation (109) will be non-positive - a crucial step for the proof of the no-short hair theorem.    
\\This trivially generalizes to the d-dimensional case as well, much like the procedure for d-dimensional general theory of relativity. 
\section{Appendix C - Radial form of the Isolated Braneworld Wormhole}
Following \cite{Biswas_2024}, we can write the metric of the on-brane wormhole (71) in radial coordinates as,
\begin{equation}
    ds^{2} = -\frac{1}{(\chi + 1)^{2}}
\left(
\chi + \sqrt{1 - \frac{2M_{1}}{r}}
\right)^{2} dt^{2}
+ \left(1 - \frac{2M_{1}}{r}\right)^{-1} dr^{2}
+ r^{2} d\Omega_{2}^{2}
\end{equation}
where,
\begin{equation}
    \Phi(r) =
\frac{\Phi_{2}}{4}
\left[
\ln\left(
\chi + \sqrt{1 - \frac{2M_{1}}{r}}
\right)
\right]^{2}
+ \Phi_{1}\sqrt{\Phi_{0}+1}\,
\ln\left(
\chi + \sqrt{1 - \frac{2M_{1}}{r}}
\right)
+ \Phi_{0}
\end{equation}
The Goldberger-Wise mechanism being held true is implicitly assumed, such that the two branes never collide. For $\chi = 0$, we get a constant radion field which is equivalent to saying that there is no observable extra-dimension before it completely collapses. The on-brane matter satisfies the trace of its anisotropic fluid EM-tensor being zero. We also note here the total EM-tensor (on-brane matter + radion field EM-tensor),
\begin{equation}
    T^{(b)\,\mu}_{\nu}
\equiv
\frac{1}{\Phi}\, T^{\mu}_{\nu}
+ \frac{\ell}{\kappa^{2}\Phi}\,
T^{\mu}_{\nu}(\Phi)
=
\mathrm{diag}\!\left(
0,\,
p^{(b)}_{r},\,
p^{(b)}_{\perp},\,
p^{(b)}_{\perp}
\right)
\end{equation}
such that,
\begin{equation}
    p^{(b)}_{r}
=
-\frac{\chi M_{1}}{8\pi r^{3}}
\left(
\chi + \sqrt{1 - \frac{2M_{1}}{r}}
\right), p^{(b)}_{\perp}
=
\frac{2\chi M_{1}}{8\pi r^{3}}
\left(
\chi + \sqrt{1 - \frac{2M_{1}}{r}}
\right)
\end{equation}
This EM-tensor is not traceless, violates both the NEC and the WEC, and can be easily checked that it is indeed a wormhole geometry by noting that the spacetime becomes non-Riemannian within $r=2M_1$. However, in our present work, we are only concerned with the on-brane Jordan frame observer and hence the satisfaction of the WEC for that is enough for us. However, it might interest one to understand how the no-short hair theorem might change for galactic wormholes where things are much better-behaved owing to dark matter halo.  
\bibliographystyle{plain}
\bibliography{references}
\end{document}